\documentclass{ejm}
\usepackage{amsbsy}

\begin{document}


\def\eqref#1{(\ref{#1})}
\def\eqrefs#1#2{(\ref{#1}) and~(\ref{#2})}
\def\eqsref#1#2{(\ref{#1}) to~(\ref{#2})}
\def\sysref#1#2{(\ref{#1}),~(\ref{#2})}

\def\Eqref#1{Eq.~(\ref{#1})}
\def\Eqrefs#1#2{Eqs.~(\ref{#1}) and~(\ref{#2})}
\def\Eqsref#1#2{Eqs.~(\ref{#1}) to~(\ref{#2})}
\def\Sysref#1#2{Eqs. (\ref{#1}),~(\ref{#2})}

\def\secref#1{Sec.~\ref{#1}}
\def\secrefs#1#2{Sec.~\ref{#1} and~\ref{#2}}

\def\appref#1{Appendix~\ref{#1}}

\def\Ref#1{Ref.~\cite{#1}}

\def\Cite#1{${\mathstrut}^{\cite{#1}}$}

\def\tableref#1{Table~\ref{#1}}

\def\figref#1{Fig.~\ref{#1}}

\hyphenation{Eq Eqs Sec App Ref Fig}

\def\EQ{\begin{equation}}
\def\EQs{\begin{eqnarray}}
\def\endEQ{\end{equation}}
\def\endEQs{\end{eqnarray}}

\def\Thm{\begin{theorem}}
\def\endThm{\end{theorem}}
\def\Lem{\begin{lemma}}
\def\endLem{\end{lemma}}
\def\Cor{\begin{corollary}}
\def\endCor{\end{corollary}}
\def\Prop{\begin{proposition}}
\def\endProp{\end{proposition}}
\def\Defn{\begin{definition}}
\def\endDefn{\end{definition}}
\def\Proof[#1]{\begin{proof}[#1]}
\def\endProof{\end{proof}}

\def\Rem[#1]{{\bf {#1}}}
\def\proclaim#1{\medbreak
\noindent{\it {#1}}\par\medbreak}


\def\fewquad{\qquad\qquad}
\def\severalquad{\qquad\fewquad}
\def\manyquad{\qquad\severalquad}
\def\manymanyquad{\manyquad\manyquad}

\def\downupindices#1#2{{}^{}_{#1}{}_{}^{#2}}
\def\updownindices#1#2{{}_{}^{#1}{}^{}_{#2}}
\def\mixedindices#1#2{{\mathstrut}^{#1}_{\mathstrut #2}}
\def\downindex#1{{\mathstrut}^{\mathstrut}_{#1}}
\def\upindex#1{{\mathstrut}_{\mathstrut}^{#1}}

\def\eqtext#1{\hbox{\rm{#1}}}

\def\Parder#1#2{
\mathchoice{\partial{#1} \over\partial{#2}}{\partial{#1}/\partial{#2}}{}{} }
\def\nParder#1#2#3{
\mathchoice{\partial^{#1}{#2} \over\partial{#3}^{#1}}{\partial^{#1}{#2}/\partial{#3}^{#1}}{}{} }
\def\mixedParder#1#2#3#4{
\mathchoice{\partial^{#1}{#2} \over\partial{#3}\partial{#4}}{\partial^{#1}{#2}/\partial{#3}\partial{#4}}{}{} }
\def\parder#1#2{\partial{#1}/\partial{#2}}
\def\nparder#1#2{\partial^{#1}/(\partial {#2})^{#1}}
\def\mixedparder#1#2#3#4{\partial^{#1}{#2}/\partial{#3}\partial{#4}}
\def\parderop#1{\partial/\partial{#1}}
\def\mixedparderop#1#2#3{\partial^{#1}/\partial{#2}\partial{#3}}

\def\x#1{x^{#1}}
\def\u#1{u\upindex{#1}}
\def\bdx{{\boldsymbol x}}
\def\bdu{{\boldsymbol u}}
\def\ujet#1{\nder{}{#1}u}
\def\bdujet#1{\nder{}{#1}\bdu}
\def\uder#1#2{u\mixedindices{#1}{#2}}
\def\deru#1#2{\der{#2}{\u{#1}}}
\def\nderu#1#2#3{\nder{#2}{#3}\u{#1}}
\def\purederu#1#2#3{\nParder{#2}{\u{#1}}{#3}}
\def\mixderu#1#2#3#4{\mixedParder{#2}{\u{#1}}{#3}{#4}}

\def\D#1{D\downindex{#1}}
\def\G#1{G\upindex{#1}}
\def\linop#1#2#3{({\cal L}_{#1})\mixedindices{#2}{#3}}
\def\adlinop#1#2#3{({\cal L}^*_{#1})\mixedindices{#2}{#3}}
\def\symm#1{\eta\upindex{#1}}
\def\adsymm#1{\omega\downindex{#1}}
\def\factor#1{\Lambda\downindex{#1}}  
\def\bdfactor{{\boldsymbol \Lambda}}
\def\Udersymm#1#2#3{\Parder{\symm{#1}}{\uder{#2}{#3}}}
\def\Uderadsymm#1#2#3{\Parder{\adsymm{#1}}{\uder{#2}{#3}}}
\def\Uderfactor#1#2#3{\Parder{\factor{#1}}{\uder{#2}{#3}}}
\def\Dfactor#1#2{\Lambda\downindex{#1#2}}

\def\factorseq#1{\factor{(#1)}}

\def\ujetl{\ujet{\ell}}

\def\g#1{g\upindex{#1}} 
\def\bdg{{\boldsymbol g}}
\def\Uderg#1#2#3{\Parder{\g{#1}}{\uder{#2}{#3}}}
\def\parderg#1#2{\Parder{\g{#1}}{{#2}}}

\def\M#1{M\downindex{#1}}
\def\P#1{\Phi\upindex{#1}}
\def\S#1{S\upindex{#1}}
\def\Phat#1{{\tilde\Phi}\upindex{#1}}  
\def\K{K}
\def\Y#1{A\upindex{#1}}
\def\T#1{T\upindex{#1}}
\def\C#1#2{C\mixedindices{#2}{#1}}

\def\w#1{w\downindex{#1}}
\def\v#1{v\upindex{#1}}
\def\bdv{{\boldsymbol v}}
\def\W#1{W\downindex{#1}}
\def\V#1{V\upindex{#1}}
\def\Wt#1{W\upindex{#1}}
\def\vder#1#2{v\mixedindices{#1}{#2}}

\def\tU#1{{\tilde U}\upindex{#1}}
\def\tu#1{{\tilde u}\upindex{#1}}  
\def\bdtu{{\tilde \bdu}}  
\def\tuder#1#2{{\tilde u}\mixedindices{#1}{#2}}
\def\uparm#1#2{\uder{#1}{(#2)}}
\def\uparmder#1#2#3{\uder{#1}{(#2)#3}}
\def\ul{u\downindex{(\lambda)}}
\def\uljet#1{\nder{}{#1}u\downindex{(\lambda)}}
\def\bdul{\bdu\downindex{(\lambda)}}
\def\bduljet#1{\nder{}{#1}\bdu\downindex{(\lambda)}}

\def\Pder#1#2{\Phi\mixedindices{#1}{#2}}

\def\X#1{{\rm X}_{#1}}
\def\ELop#1#2{E_{#1\upindex{#2}}}
\def\parderU#1#2{\der{\uder{#1}{#2}}}
\def\ELophat#1#2{{\hat E}_{#1\upindex{#2}}} 
\def\elop#1{E_{#1}}
\def\elophat#1{{\hat E}_{#1}}

\def\nD#1#2{D\mixedindices{#1}{#2}}
\def\der#1{\partial_{#1}}
\def\nder#1#2{\partial^{#2}_{#1}}
\def\Dt#1{D\mixedindices{#1}{t}}
\def\Dx#1{D\mixedindices{#1}{x}}
\def\solD#1{{\cal D}\downindex{#1}}
\def\solnD#1#2{{\cal D}\mixedindices{#1}{#2}}

\def\Du#1{u_{#1}}
\def\bdDu#1{{\bdu}_{#1}}
\def\Dtu#1{{\tilde u}_{#1}}
\def\bdDtu#1{{\tilde \bdu}_{#1}}  
\def\bdDul#1{\bdu\downindex{(\lambda){#1}}}
\def\Dv#1{v_{#1}}

\def\slinop#1{{\cal L}_{#1}}
\def\adslinop#1{{\cal L}^*_{#1}}

\def\N{N}
\def\measure{{d\lambda \over\lambda}}
\def\tc{\tilde c}

\def\h{h}

\def\pde/{partial differential equation}
\def\de/{differential equation}
\def\conslaw/{conservation law}
\def\adsys/{adjoint system}
\def\adinv/{adjoint invariance condition}
\def\hsys/{extra system}
\def\dcm/{direct conservation law method}
\def\CK/{Cauchy-Kovalev\-skaya}

\def\ie/{i.e.}
\def\eg/{e.g.}
\def\etc/{etc.}
\def\const{{\rm const}}


\def\pot{\psi}
\def\Dpot#1{\psi_{#1}}
\def\Pot{\psi}
\def\DPot#1{\psi_{#1}}
\def\potsymm{\varrho}
\def\potsymmseq#1{\varrho_{(#1)}}
\def\wder#1{\potsymm\downindex{#1}}

\def\linG{{\cal L}_G}
\def\adlinG{{\cal L}^*_G}
\def\linl{{\cal L}_\l}
\def\adlinl{{\cal L}^*_\l}
\def\lin#1{{\cal L}_{#1}}
\def\adlin#1{{\cal L}^*_{#1}}   

\def\invDx{D\mixedindices{-1}{x}}

\def\symmseq#1{\eta_{(#1)}}
\def\adsymmseq#1{\omega_{(#1)}} 

\def\l{\Lambda}
\def\lhat{{\hat\Lambda}}
\def\lder#1{\Lambda\downindex{#1}}
\def\adsymmder#1{\omega\downindex{#1}} 
\def\adsymmseqder#1#2{\omega_{(#1)#2}} 

\def\lparmu{\lambda u}


\def\Dhat#1{{\hat D}\downindex{#1}}

\newtheorem{theorem}{Theorem}[section]
\newtheorem{corollary}[theorem]{Corollary}
\newtheorem{lemma}[theorem]{Lemma}
\newtheorem{proposition}[theorem]{Proposition}

\newdefinition{definition}[theorem]{Definition}


\title[Direct conservation law method]
{ Direct construction method for conservation \\
laws of partial differential equations. Part II: \\
General treatment }

\author[ S.\ns C.\ns A\ls N\ls C\ls O\ns
\and G.\ns B\ls L\ls U\ls M\ls N\ls A\ls N]
{ S\ls T\ls E\ls P\ls H\ls E\ls N\ns C.\ns A\ls N\ls C\ls O$\,^1$\ns
\and G\ls E\ls O\ls R\ls G\ls E\ns B\ls L\ls U\ls M\ls A\ls N$\,^2$ }
\affiliation{
$^1\,$Department of Mathematics, 
Brock University,
St. Catharines, ON Canada L2S 3A1\\
email\textup{\nocorr: \texttt{sanco@brocku.ca}}\\
$^2\,$Department of Mathematics, 
University of British Columbia, 
Vancouver, BC Canada V6T 1Z2 \\
email\textup{\nocorr: \texttt{bluman@math.ubc.ca}} }

\maketitle

\begin{abstract}
This paper gives a general treatment and proof of
the direct conservation law method presented in Part I
(see \cite{partI}). 
In particular, 
the treatment here applies to finding the local conservation laws of 
any system of one or more partial differential equations
expressed in a standard \CK/ form. 
A summary of the general method 
and its effective computational implementation is also given. 
\end{abstract}

\section{Introduction}
\label{introII}

In this paper 
we present a general treatment of the \dcm/
introduced in Part I (see \Ref{partI}). 
In particular, in \secref{method}
we show how to find the local conservation laws
for any system of one or more PDEs
expressed in a standard \CK/ form.
We specifically treat $n$th order scalar PDEs
in \secref{scalarcase}. 
In \secref{conclude}
we summarize the general method
and discuss its effective implementation in computational terms. 

In order to make the treatment uniform, 
it is convenient to work with 
\CK/ systems of PDEs as follows. 

\Defn\label{defnCK}
A PDE system with any number of independent and dependent variables
has {\it \CK/ form} in terms of a given independent variable
if the system is in solved form for 
a pure derivative of the dependent variables 
with respect to the given independent variable,
and if all other derivatives of dependent variables in the system 
are of lower order with respect to that independent variable.
\endDefn

Typically, scalar PDEs admit a \CK/ form
by singling out a derivative with respect to one independent variable,
or by making a point transformation 
(more generally a contact transformation) 
on the independent variables. 
For example:
the wave equation 
\EQs
\Du{tx}=0
\nonumber
\endEQs
admits the \CK/ form $\Du{tt}=\Du{xx}$
after the point transformation $t\rightarrow t-x$, $x\rightarrow x+t$;
the harmonic equation 
\EQs
\Du{xx}+\Du{yy}=0
\nonumber
\endEQs
admits the \CK/ form $\Du{yy}=-\Du{xx}$ 
with respect to $y$. 
A less trivial example is the Kadomtsev-Petviashvili equation \cite{kpeq}
\EQs
\Du{tx}+(u\Du{x})_x +\Du{xxxx} \pm \Du{yy}=0 . 
\nonumber
\endEQs
This equation admits two obvious \CK/ forms:
$\Du{yy}= \mp( \Du{tx}+(u\Du{x})_x +\Du{xxxx} )$
which is a second-order PDE with respect to $y$;
and $\Du{xxxx}= \mp \Du{yy} -\Du{tx}-(u\Du{x})_x$
which is a fourth-order PDE with respect to $x$.

As examples which are more involved, consider 
the modified Benjamin-Bona-Mahoney equation \cite{mbbmeq}
\EQs
\Du{t}+(1+u^2)\Du{x} -\Du{xxt}=0 , 
\nonumber
\endEQs
and the symmetric regularized long wave equation \cite{srlweq}
\EQs
\Du{tt} +\Du{xx}+u\Du{tx} +\Du{x}\Du{t} +\Du{ttxx}=0 . 
\nonumber
\endEQs
As it stands the modified Benjamin-Bona-Mahoney equation 
is not of \CK/ form with respect to either $t$ or $x$, 
since the $t$ derivatives of $u$ appear in both 
pure and mixed derivative terms, 
while the highest order $x$ derivative of $u$ appears in a mixed derivative
involving $t$ and hence is not in solved form. 
Nevertheless, if one makes the point transformation 
$t\rightarrow t$, $x\rightarrow x-t$,
then the modified Benjamin-Bona-Mahoney equation becomes 
$\Du{xxx}-\Du{xxt}+u^2\Du{x} +\Du{t}=0$
which now is of third-order \CK/ form with respect to $x$. 
The situation for the symmetric regularized long wave equation is similar. 
It is not of \CK/ form as it stands, 
but after one makes the point transformation 
$t\rightarrow t-x$, $x\rightarrow x+t$
it is of fourth-order \CK/ form with respect to $t$ or $x$:
$\Du{tttt}+\Du{xxxx} -2\Du{ttxx} +(2-u)\Du{tt}+(2+u)\Du{xx}
+\Du{t}{}^2 -\Du{x}{}^2=0$.

Many PDE systems can be handled similarly to scalar PDEs.
For example, 
the vector nonlinear Schroedinger equation 
\EQs
i\vec{u}_t + \vec{u}_{xx} \pm  f(|\vec{u}|)\vec{u} =0, 
\quad 
\vec{u}=(\u{1},\ldots,\u{n})
\nonumber
\endEQs
admits the first-order \CK/ form 
$\vec{u}_t = i\vec{u}_{xx} \pm  if(|\vec{u}|) \vec{u}$
with respect to $t$,
as well as the second-order \CK/ form 
$\vec{u}_{xx} =-i\vec{u}_t \mp f(|\vec{u}|) \vec{u}$
with respect to $x$. 
A less obvious example is Navier's equations of isotropic elasticity,
\EQs
&& \kappa\Du{xx} + \mu \Du{yy} + (\kappa-\mu) \Dv{xy} =0, 
\nonumber\\
&& (\kappa-\mu) \Du{xy} + \mu\Dv{xx} + \kappa \Dv{yy} =0,
\nonumber
\endEQs
$\kappa=\const,\mu=\const$.
This PDE system admits a second-order \CK/ form with respect to $x$ or $y$:
$\Du{xx} = -\frac{\mu}{\kappa} \Du{yy} +(\frac{\mu}{\kappa}-1) \Dv{xy}$
and 
$\Dv{xx} = -\frac{\kappa}{\mu} \Dv{yy} +(1-\frac{\kappa}{\mu}) \Du{xy}$.

In general any \CK/ form of a system of one or more PDEs
can be used with no loss of completeness 
in finding the \conslaw/s admitted by the system.
Given a \CK/ PDE system, 
we let $t$ denote the independent variable
in the derivative which appears in solved form in the PDEs,
with the remaining independent variables denoted by 
$\bdx=(\x{1},\ldots,\x{n})$. 
In order to obtain the most effective formulation of the \dcm/, 
it is convenient to express the system 
in its equivalent first-order (evolution) form with respect to $t$. 

Hence, we consider a first-order \CK/ system of PDEs 
with $N$ dependent variables $\bdu=(\u{1},\ldots,\u{N})$
and $n+1$ independent variables $(t,\bdx)$,
\EQ\label{uPDE}
\G{\sigma} = 
\Parder{\u{\sigma}}{t} 
+ \g{\sigma}(t,\bdx,\bdu,\der{\bdx}\bdu,\ldots,\nder{\bdx}{m}\bdu)
=0 ,\ \sigma=1,\ldots,N
\endEQ
with $\bdx$ derivatives of $\bdu$ up to some order $m$. 
We use $\der{\bdx}\bdu,\nder{\bdx}{2}\bdu,$ \etc/ 
to denote all derivatives of $\u{\sigma}$ 
of a given order with respect to $\x{i}$. 
We denote partial derivatives 
$\partial/\partial t$ and $\partial/\partial \x{i}$
by subscripts $t$ and $i$ respectively.
Corresponding total derivatives are denoted by
$\D{t}$ and $\D{i}$.
We let $\linop{g}{\sigma}{\rho}$ 
denote the linearization operator of $\g{\sigma}$ defined by
\EQ\label{ling}
\linop{g}{\sigma}{\rho} \V{\rho} =
\Uderg{\sigma}{\rho}{} \V{\rho} 
+ \Uderg{\sigma}{\rho}{i} \D{i}\V{\rho} 
+\cdots
+ \Uderg{\sigma}{\rho}{i_1\cdots i_m} \D{i_1\cdots i_m}\V{\rho} , 
\endEQ
and we let $\adlinop{g}{\sigma}{\rho}$ 
denote the adjoint operator defined by
\EQ\label{adling}
\adlinop{g}{\sigma}{\rho} \W{\sigma} =
\Uderg{\sigma}{\rho}{} \W{\sigma} 
-\D{i}( \Uderg{\sigma}{\rho}{i} \W{\sigma} )
+\cdots
+ (-1)^m \D{i_1\cdots i_m}( 
\Uderg{\sigma}{\rho}{i_1\cdots i_m} \W{\sigma} ) , 
\endEQ
acting on arbitrary functions $\V{\rho},\W{\sigma}$ respectively.

Throughout we use the summation convention for 
{\it repeated} lower-case indices; 
we use an explicit summation sign where needed for summing over non-indices.

\section{General treatment}
\label{method}

We start by considering the determining equations for
symmetries and adjoint symmetries.
Suppose $\X{}$ is the infinitesimal generator of a symmetry
leaving invariant PDE system \eqref{uPDE}.
We denote $\X{}\u{\sigma} = \symm{\sigma}$, which satisfies
\EQ\label{symmeq}
0=\D{t}\symm{\sigma} +\linop{g}{\sigma}{\rho} \symm{\rho},\
\sigma=1,\ldots,N
\endEQ
for all solutions $\bdu(t,\bdx)$ of \Eqref{uPDE}.
This linearization of \Eqref{uPDE} 
is the determining equation for symmetries 
(point-type as well as first-order and higher-order type \cite{olverbook})
$\symm{\sigma}(t,\bdx,\bdu,\bdujet{},\ldots,\bdujet{p})$
of the PDE system \eqref{uPDE},
where $\bdujet{j}$ denotes all $j$th order derivatives of $\bdu$
with respect to all independent variables $t,\bdx$.
The adjoint of \Eqref{symmeq} is given by 
\EQ\label{adsymmeq}
0=-\D{t}\adsymm{\sigma} +\adlinop{g}{\rho}{\sigma} \adsymm{\rho},\
\sigma=1,\ldots,N
\endEQ
which is the determining equation for adjoint symmetries
$\adsymm{\sigma}(t,\bdx,\bdu,\bdujet{},\ldots,\bdujet{p})$
of the PDE system \eqref{uPDE}.
In general, 
solutions of the adjoint symmetry equation \eqref{adsymmeq}
are not solutions of the symmetry equation \eqref{symmeq},
and there is no interpretation of adjoint symmetries in terms of
an infinitesimal generator leaving anything invariant. 

In order to solve the determining equations for 
$\symm{\sigma}$ and $\adsymm{\sigma}$,
one works on the space of solutions of the PDE system. 
This means we use the PDEs to eliminate $\uder{\sigma}{t}$
in terms of $\u{\sigma}$, $\uder{\sigma}{i}$, \etc/ 
In particular, without loss of generality, 
we are free to let $\symm{\sigma}$ and $\adsymm{\sigma}$ 
have no dependence on $\uder{\sigma}{t}$ and its differential consequences.
Let
\EQ
\solD{t} = 
\der{t} 
-( \g{\rho}\der{\u{\rho}} +(\D{i}\g{\rho})\der{\uder{\rho}{i}} +\cdots )
\endEQ
which is the total derivative with respect to $t$ 
on the solution space of PDE system \eqref{uPDE}.
(In particular, $\D{t}=\solD{t}$
when acting on all solutions $\bdu(t,\bdx)$.)
Then the determining equations explicitly become
\EQs
0 = &&
\solD{t}\symm{\sigma} +\linop{g}{\sigma}{\rho} \symm{\rho} 
\nonumber\\
= &&
\Parder{\symm{\sigma}}{t} 
-\bigg( \Udersymm{\sigma}{\rho}{} \g{\rho}
+\Udersymm{\sigma}{\rho}{i} \D{i}\g{\rho}
+\cdots
+\Udersymm{\sigma}{\rho}{i_1\cdots i_p} \D{i_1}\cdots\D{i_p}\g{\rho} 
\bigg)
\nonumber\\&&\quad
+ \Uderg{\sigma}{\rho}{} \symm{\rho} 
+ \Uderg{\sigma}{\rho}{i} \D{i}\symm{\rho}
+\cdots
+\Uderg{\sigma}{\rho}{i_1\cdots i_m} \D{i_1}\cdots \D{i_m}\symm{\rho},\
\sigma=1,\ldots,N
\label{jetsymmeq}
\endEQs
for $\symm{\sigma}(t,\bdx,\bdu,\der{\bdx}\bdu,\ldots,\nder{\bdx}{p}\bdu)$,
and
\EQs
0 = &&
-\solD{t}\adsymm{\sigma} +\adlinop{g}{\rho}{\sigma} \adsymm{\rho} 
\nonumber\\
= &&
-\Parder{\adsymm{\sigma}}{t}
+\bigg( \Uderadsymm{\sigma}{\rho}{} \g{\rho}
+\Uderadsymm{\sigma}{\rho}{i} \D{i}\g{\rho}
+\cdots
+\Uderadsymm{\sigma}{\rho}{i_1\cdots i_p} \D{i_1}\cdots\D{i_p}\g{\rho}
\bigg)
\nonumber\\&&\quad
+ \Uderg{\rho}{\sigma}{} \adsymm{\rho}
-\D{i}\Big( \Uderg{\rho}{\sigma}{i} \adsymm{\rho} \Big)
+\cdots
+(-1)^m \D{i_1}\cdots \D{i_m}\Big(
\Uderg{\rho}{\sigma}{i_1\cdots i_m} \adsymm{\rho} \Big),\
\sigma=1,\ldots,N
\nonumber\\
\label{jetadsymmeq}
\endEQs
for $\adsymm{\sigma}(t,\bdx,\bdu,\der{\bdx}\bdu,\ldots,\nder{\bdx}{p}\bdu)$.
The solutions of \Eqrefs{jetsymmeq}{jetadsymmeq}
yield all symmetries and adjoint symmetries
up to any given order $p$.

We now consider \conslaw/s.
\Defn\label{defnconslaw}
A {\it local \conslaw/} of PDE system \eqref{uPDE} 
is a divergence expression
\EQ\label{ucons}
\D{t}\P{t}(t,\bdx,\bdu,\bdujet{},\ldots,\bdujet{k}) 
+ \D{i}\P{i}(t,\bdx,\bdu,\bdujet{},\ldots,\bdujet{k})  = 0
\endEQ
for all solutions $\bdu(t,\bdx)$ of \Eqref{uPDE};
$\P{t}$ and $\P{i}$ are called the conserved densities. 
\endDefn

The conservation equation \eqref{ucons} holds as an identity if,
for all solutions $\bdu(t,\bdx)$ of \Eqref{uPDE},
\EQ\label{trivialcons}
\P{t}=\D{i} \theta^i, 
\P{i}=-\D{t} \theta^i +\D{j} \psi^{ij}
\endEQ
for some expressions 
$\theta^i(t,\bdx,\bdu,\bdujet{},\ldots,\bdujet{k-1})$,
$\psi^{ij}(t,\bdx,\bdu,\bdujet{},\ldots,\bdujet{k-1})$
with $\psi^{ij}=-\psi^{ji}$. 
Such \conslaw/s are trivial.
Only the nontrivial \conslaw/s of the PDE system \eqref{uPDE}
are of interest.

\Defn\label{defnnontrivial}
A local conservation law \eqref{ucons} is {\it nontrivial} iff
the conserved densities do not satisfy \Eqref{trivialcons}. 
\endDefn
Any nontrivial conserved densities that agree 
to within trivial conserved densities 
are regarded as defining the same nontrivial \conslaw/.
There is further freedom in the form of conserved densities 
since we are clearly free to replace $\uder{\sigma}{t}=-\g{\sigma}$
in $\P{t}$ and $\P{i}$ on the solution space of PDE system \eqref{uPDE}.
Thus, without loss of generality we can consider
$\P{t}$ and $\P{i}$ to depend only on $t,\bdx,\bdu$, 
and $\bdx$ derivatives of $\bdu$.
We refer to this as the {\it normal form} of the \conslaw/,
\EQ\label{normalcons}
\D{t}\P{t}(t,\bdx,\bdu,\der{\bdx}\bdu,\ldots,\nder{\bdx}{k}\bdu)
+ \D{i}\P{i}(t,\bdx,\bdu,\der{\bdx}\bdu,\ldots,\nder{\bdx}{k}\bdu) = 0  
\endEQ
for all solutions $\bdu(t,\bdx)$ of PDE system \eqref{uPDE}.
In normal form, the freedom corresponding to trivial conserved densities 
is given by 
\EQ\label{freedom}
\P{t} \rightarrow \P{t} +\D{i}\theta^i,
\P{i} \rightarrow \P{i} -\solD{t}\theta^i +\D{j}\psi^{ij}
\endEQ
where $\theta^i,\psi^{ij}=-\psi^{ji}$
do not depend on $\bdDu{t}$ and differential consequences.

All nontrivial local \conslaw/s (in normal form) 
can be shown to arise from
multipliers on the PDEs \eqref{uPDE} 
as follows. 
We move off the solution space of \Eqref{uPDE}
and let $\bdu(t,\bdx)$ be an arbitrary function of $t,\bdx$.

\Defn\label{defnmult}
{\it Multipliers} for PDE system \eqref{uPDE} 
are a set of expressions 
\EQs\nonumber
\{ \factor{1}(t,\bdx,\bdu,\bdujet{},\ldots,\bdujet{q}), \ldots,
\factor{N}(t,\bdx,\bdu,\bdujet{},\ldots,\bdujet{q}) \}
\endEQs
satisfying 
\EQ\label{multipliereq}
(\uder{\sigma}{t} +\g{\sigma}) \factor{\sigma} = 
\D{t} \Phat{t} + \D{i} \Phat{i}
\endEQ
for some expressions 
$\Phat{t}(t,\bdx,\bdu,\bdujet{},\ldots,\bdujet{k})$
and $\Phat{i}(t,\bdx,\bdu,\bdujet{},\ldots,\bdujet{k})$
for all functions $\bdu(t,\bdx)$. 
\endDefn

Given a \conslaw/ \eqref{normalcons},
consider $\D{t} \P{t} +\D{i}\P{i}$. 
Clearly this expression must be proportional to 
$\uder{\sigma}{t}+\g{\sigma}$ 
and its differential consequences
in order to satisfy \Eqref{normalcons}.
The $\uder{\sigma}{t}$ terms arise only from
\EQ
\D{t}\P{t} = 
\Parder{\P{t}}{t} + \Parder{\P{t}}{\u{\sigma}} \uder{\sigma}{t}
+ \Parder{\P{t}}{\uder{\sigma}{i}} \uder{\sigma}{ti}
+\cdots
+\Parder{\P{t}}{\uder{\sigma}{i_1\cdots i_k}} \uder{\sigma}{ti_1\cdots i_k} 
= \der{t}\P{t} + \linop{\P{t}}{}{\sigma} \uder{\sigma}{t}
\endEQ
where $\linop{\P{t}}{}{\sigma}= 
( \Parder{\P{t}}{\u{\sigma}} )
+ ( \Parder{\P{t}}{\uder{\sigma}{i}} )\D{i} 
+\cdots
+( \Parder{\P{t}}{\uder{\sigma}{i_1\cdots i_k}} )\D{i_1}\cdots\D{i_k}$
denotes the linearization operator of $\P{t}$.
To organize these terms we use the identities
\EQs
\linop{\P{t}}{}{\sigma} \uder{\sigma}{t} &&
=\linop{\P{t}}{}{\sigma} (\uder{\sigma}{t} +\g{\sigma} )
-\linop{\P{t}}{}{\sigma} \g{\sigma}
\nonumber\\&&
=  (\uder{\sigma}{t} +\g{\sigma} )\ELophat{u}{\sigma} (\P{t})
-\linop{\P{t}}{}{\sigma} \g{\sigma} +\D{i} \Gamma^i
\label{linPtids}
\endEQs
where 
$\Gamma^i$ is given by an expression proportional to 
$\uder{\sigma}{t}+\g{\sigma}$ (and differential consequences),
and where
\EQ
\ELophat{u}{\sigma} = 
\parderU{\sigma}{} -\D{i}\parderU{\sigma}{i} 
+\D{i}\D{j}\parderU{\sigma}{ij} +\cdots
\endEQ
is a restricted Euler operator. 
Thus, we have 
\EQ\label{Ptid}
\D{t}\P{t} = 
\der{t}\P{t} -\linop{\P{t}}{}{\sigma} \g{\sigma} +\D{i} \Gamma^i
+ (\uder{\sigma}{t} +\g{\sigma} )\ELophat{u}{\sigma} (\P{t}) . 
\endEQ
In order for the conservation equation \eqref{normalcons} to hold,
the terms $\der{t}\P{t} -\linop{\P{t}}{}{\sigma} \g{\sigma}$
which do not involve $\uder{\sigma}{t}+\g{\sigma}$ 
must cancel $\D{i}\P{i}$, 
and therefore we have
\EQ\label{Piid}
\D{i}\P{i} = -\der{t}\P{t} +\linop{\P{t}}{}{\sigma} \g{\sigma} . 
\endEQ
Then combining expressions \eqrefs{Ptid}{Piid}
we obtain
\EQ\label{conseq}
\D{t}\P{t} +\D{i}(\P{i}-\Gamma^i) =
( \uder{\sigma}{t} +\g{\sigma} ) \factor{\sigma}
\endEQ
with 
\EQ\label{factoreq}
\factor{\sigma} = \ELophat{u}{\sigma} (\P{t}),\
\sigma=1,\ldots,N . 
\endEQ
When $\bdu(t,\bdx)$ is restricted to 
the solution space of PDE system \eqref{uPDE},
then $\Gamma^i$ vanishes 
and the divergence expression \eqref{conseq} reduces to 
the conservation equation \eqref{normalcons}.

Hence, the expressions $\{ \ELophat{u}{\sigma} (\P{t}) \}$
define multipliers $\{ \factor{\sigma} \}$ 
yielding a \conslaw/ \eqref{normalcons}.
Furthermore, since $\P{t}$ does not depend on $\bdDu{t}$
and its differential consequences,
we see that each multiplier expression $\factor{\sigma}$ 
is a function only of $t,\bdx,\bdu$, and $\bdx$ derivatives of $\bdu$. 
Most important, these expressions $\factor{\sigma}$ 
are invariant under a change in $\P{t}$ by
a trivial conserved density \eqref{freedom}
since $\ELophat{u}{\sigma}$ annihilates divergences $\D{i}\theta^i$
where $\theta^i$ depends on $t,\bdx,\bdu$ and $\bdx$ derivatives of $\bdu$.
(In particular, if $\P{t}$ in normal form is trivial, 
then $\factor{\sigma}$ is identically zero, 
and conversely.)
Thus we have the following result. 

\Thm\label{thmmult}
For the \CK/ PDE system \eqref{uPDE}, 
{\rm every} nontrivial \conslaw/ in normal form \eqref{normalcons} 
is uniquely characterized by a set of multipliers $\{ \factor{\sigma} \}$ 
with no dependence on $\bdDu{t}$ and differential consequences,
satisfying the relations \eqrefs{conseq}{factoreq} 
holding for all functions $\bdu(t,\bdx)$.
\endThm
From this result it is natural to define 
the {\it order} of a \conslaw/ \eqref{normalcons} 
as the order of the highest $\bdx$ derivative of $\bdu$ 
in its multipliers \eqref{factoreq}.

Theorem~\ref{thmmult} is the starting point for an effective approach to 
find \conslaw/s of PDE system \eqref{uPDE} by use of multipliers.
The standard determining condition \cite{olverbook}
for multiplier expressions 
$\factor{\sigma}(t,\bdx,\bdu,\der{\bdx}\bdu,\ldots,\nder{\bdx}{p}\bdu)$
arises from the definition \eqref{multipliereq}
by the well-known result that 
divergence expressions are characterized by annihilation under
the full Euler operator
\EQ\label{elop}
\ELop{u}{\sigma} = 
\parderU{\sigma}{} -\D{i}\parderU{\sigma}{i} -\D{t}\parderU{\sigma}{t} 
+\D{i}\D{j}\parderU{\sigma}{ij} 
+\D{t}\D{j}\parderU{\sigma}{tj} 
+\cdots . 
\endEQ
This yields (by a straightforward calculation)
\EQ\label{multiplierdeteq}
0=\ELop{u}{\sigma} (\uder{\rho}{t}\factor{\rho} + \g{\rho}\factor{\rho})
= -\D{t}\factor{\sigma} +\adlinop{g}{\rho}{\sigma}\factor{\rho}
+\adlinop{\factor{}}{}{\sigma\rho} (\uder{\rho}{t} +\g{\rho}),\
\sigma=1,\ldots,N
\endEQ
where $\adlinop{\factor{}}{}{\sigma\rho}$ is the adjoint operator of
the linearization operator $\linop{\factor{}}{}{\sigma\rho}$
defined by
\EQ
\linop{\factor{}}{}{\sigma\rho} \V{\rho} =
\Uderfactor{\sigma}{\rho}{} \V{\rho}
+\Uderfactor{\sigma}{\rho}{i} \D{i}\V{\rho}
+\cdots
+\Uderfactor{\sigma}{\rho}{i_1\cdots i_p} \D{i_1}\cdots\D{i_p}\V{\rho}
\endEQ
and
\EQ
\adlinop{\factor{}}{}{\rho\sigma} \Wt{\sigma}=
\Uderfactor{\sigma}{\rho}{} \Wt{\sigma}
-\D{i}( \Uderfactor{\sigma}{\rho}{i} \Wt{\sigma} )
+\cdots
+(-1)^p \D{i_1}\cdots\D{i_p}(
\Uderfactor{\sigma}{\rho}{i_1\cdots i_p} \Wt{\sigma} )
\endEQ
acting on arbitrary functions $\V{\rho},\Wt{\sigma}$.
Here the determining condition \eqref{multiplierdeteq}
is required to hold for all functions $\bdu(t,\bdx)$,
\ie/, this is necessary and sufficient for 
$\uder{\rho}{t}\factor{\rho} + \g{\rho}\factor{\rho}$
to be a divergence expression.
We give a simple direct proof in \secref{proofs}.

We now show how to convert the determining condition for $\factor{\sigma}$
into a system of determining equations
that allow one to work entirely on the space of solutions of
PDE system \eqref{uPDE} to find $\factor{\sigma}$. 
Furthermore, we show that the resulting determining system
consists of the adjoint symmetry determining equation \eqref{jetadsymmeq}
augmented by extra determining equations 
giving necessary and sufficient conditions 
for an adjoint symmetry to be a set of multipliers
yielding a \conslaw/.

\subsection{Conservation law determining system}
\label{multiplierdetsys}
 
In the determining condition \eqref{multiplierdeteq}
for $\factor{\sigma}(t,\bdx,\bdu,\der{\bdx}\bdu,\ldots,\nder{\bdx}{p}\bdu)$ 
of order $p$ 
consider the terms involving $\uder{\sigma}{t}$. 
These terms arise just from 
$\D{t}\factor{\sigma}$
and $\adlinop{\factor{}}{}{\sigma\rho} \uder{\rho}{t}$,
and so it follows that 
\Eqref{multiplierdeteq} is a linear polynomial in 
$\uder{\sigma}{t}$ 
and differential consequences of $\uder{\sigma}{t}$ with respect to $\bdx$.
Since $\u{\sigma}$ is required to be an arbitrary function of 
$t$ and $\bdx$,
\Eqref{multiplierdeteq} splits into separate equations
given by the coefficients of $\uder{\sigma}{t}$, $\uder{\sigma}{ti}$, \etc/
It is convenient to organize this splitting in terms of 
$\uder{\sigma}{t}+\g{\sigma}=\G{\sigma}$ and differential consequences
$\uder{\sigma}{ti}+ \D{i}\g{\sigma}= \D{i}\G{\sigma}$, \etc/,
which we refer to as the {\it leading terms}
(all other terms in the splitting are then referred to as {\it non-leading}).
Then the leading and non-leading terms in the splitting 
must vanish separately.

To carry out the splitting of $\D{t}\factor{\sigma}$,
we use the identity
\EQs
\D{t} = 
\solD{t} 
+ (\uder{\rho}{t}+\g{\rho}) \der{\u{\rho}}
+(\uder{\rho}{ti}+\D{i}\g{\rho}) \der{\uder{\rho}{i}}
+\cdots
\nonumber
\endEQs
which yields 
$\D{t}\factor{\sigma} = 
\solD{t}\factor{\sigma} +\linop{\factor{}}{}{\sigma\rho} \G{\rho}$.

Consequently, the non-leading terms in \Eqref{multiplierdeteq} are given by
\EQs
0 =&&
-\solD{t} \factor{\sigma} + \adlinop{\g{}}{\rho}{\sigma} \factor{\rho}
\nonumber\\
=&&
-\Parder{\factor{\sigma}}{t}
+\bigg( \Uderfactor{\sigma}{\rho}{} \g{\rho}
+\Uderfactor{\sigma}{\rho}{i} \D{i}\g{\rho}
+\cdots
+\Uderfactor{\sigma}{\rho}{i_1\cdots i_p} \D{i_1}\cdots\D{i_p}\g{\rho}
\bigg)
\nonumber\\&&\fewquad
+ \Uderg{\rho}{\sigma}{} \factor{\rho}
- \D{i}\Big( \Uderg{\rho}{\sigma}{i} \factor{\rho} \Big)
+\cdots
+ (-1)^m \D{i_1}\cdots \D{i_m}\Big(
\Uderg{\rho}{\sigma}{i_1\cdots i_m} \factor{\rho} \Big) , 
\nonumber\\&&
\sigma=1,\ldots,N . 
\label{adsymmsys}
\endEQs             
This is the adjoint symmetry equation \eqref{jetadsymmeq}
with $\adsymm{\sigma}=\factor{\sigma}$. 

The leading terms in \Eqref{multiplierdeteq} are given by 
\EQ\label{adinv}
0=
-\linop{\factor{}}{}{\sigma\rho} \G{\rho}
+ \adlinop{\factor{}}{}{\sigma\rho} \G{\rho},\
\sigma=1,\ldots,N . 
\endEQ
which we call the {\it adjoint invariance condition} on $\factor{\sigma}$.
Now since $\u{\sigma}$ is required to be an arbitrary function of 
$t$ and $\bdx$,
we observe that 
\Eqref{adinv} splits into separate equations
given by the coefficients of 
$\G{\sigma}$, $\D{i}\G{\sigma}$,$\ldots$,$\D{i_1}\cdots\D{i_p}\G{\sigma}$:
\EQs
&& 0= (-1)^{p+1} \Uderfactor{\sigma}{\rho}{i_1\cdots i_p}
+\Uderfactor{\rho}{\sigma}{i_1\cdots i_p} , 
\nonumber\\
&& 0= (-1)^{q+1} \Uderfactor{\sigma}{\rho}{i_1\cdots i_q}
+ \Uderfactor{\rho}{\sigma}{i_1\cdots i_q}
-\C{q}{q+1}\D{i_{q+1}} \Uderfactor{\rho}{\sigma}{i_1\cdots i_{q+1}}
+ \cdots
\nonumber\\&&\fewquad
+ (-1)^{p-q} \C{q}{p} \D{i_{q+1}}\cdots\D{i_p}
\Uderfactor{\rho}{\sigma}{i_1\cdots i_p} , \quad
q=1,\ldots,p-1
\nonumber\\
&& 0= -\Uderfactor{\sigma}{\rho}{}
+ \Uderfactor{\rho}{\sigma}{}
-\D{i} \Uderfactor{\rho}{\sigma}{i} 
+ \cdots
+ (-1)^p \D{i_1}\cdots\D{i_p} \Uderfactor{\rho}{\sigma}{i_1\cdots i_p} , 
\label{helmholtzsysc}
\nonumber\\&&
\sigma=1,\ldots,N; \rho=1,\ldots,N
\label{splitsys}
\endEQs      
where $\C{q}{r}=\frac{r!}{q!(r-q)!}$.
This establishes the following important splitting result. 

\Lem\label{lemsplit}
For $\factor{\sigma}$ 
with no dependence on $\bdDu{t}$ and differential consequences, 
the Euler operator equation \eqref{multiplierdeteq}
is equivalent to the split system of equations
\eqrefs{adsymmsys}{splitsys},
which are required to hold for all functions $\bdu(t,\bdx)$. 
\endLem

Consequently, 
by combining Lemma~\ref{lemsplit} and Theorem~\ref{thmmult}, 
we see that \Eqrefs{adsymmsys}{splitsys}
constitute a necessary and sufficient
determining system for finding multipliers $\{\factor{\sigma}\}$. 
The number of equations in this system is 
$\frac{N^2 (n+p-1)!}{n!(p-1)!}
+\frac{N(N-(-1)^p)}{2}\frac{(n+p)!}{n!p!}$.

\Thm\label{thmdeteq}
For the \CK/ PDE system \eqref{uPDE},
the multipliers for {\rm all} nontrivial \conslaw/s 
in normal form \eqref{normalcons}
up to any given order $p$
are the solutions 
$\factor{\sigma}(t,\bdx,\bdu,\der{\bdx}\bdu,\ldots,\nder{\bdx}{p}\bdu)$
of the determining system consisting of
the adjoint symmetry determining equation \eqref{adsymmsys} 
augmented by the extra determining equations \eqref{splitsys}.
In particular, 
\Eqref{splitsys} gives necessary and sufficient conditions 
for an adjoint symmetry to be a set of multipliers. 
\endThm

In deriving the determining system for $\factor{\sigma}$,
we have eliminated $\bdDu{t}$ and its differential consequences. 
As a result, one is able to work equivalently on 
the space of solutions of the PDE system \eqref{uPDE}
in order to solve the determining system to find $\factor{\sigma}$.
In particular, 
the same algorithmic procedures which one uses to solve 
determining equations for symmetries 
can be used to solve the determining system for multipliers. 
Moreover, there is freedom in mixing the order of solving 
the determining equations in this system. 
A direct (naive) approach is 
to solve the adjoint symmetry determining equation first,
then check which of these adjoint symmetries satisfy 
the extra determining equations.
As illustrated in the examples in Part I, 
a more effective approach 
is to use the extra determining equations first.

\Rem[Remarks on the extra determining equations:]

There is a simple interpretation of 
the extra determining equations \eqref{splitsys}.
From relation \eqref{factoreq} between multipliers and conserved densities,
we observe that $\factor{\sigma}$ is a variational expression
(\ie/ it arises as an Euler-Lagrange expression from $\P{t}$).
The well-known necessary and sufficient (Helmholtz) conditions 
\cite{olverbook}
for an expression to be variational are that 
its linearization operator is self-adjoint,
and thus $\factor{\sigma}$ is a variational expression 
if and only if it satisfies \cite{fokas,fuchssteiner}
\EQ\label{selfadjointfactor}
\linop{\factor{}}{}{\sigma\rho} =\adlinop{\factor{}}{}{\sigma\rho} ,\
\sigma, \rho=1,\ldots,N . 
\endEQ
The operator equation \eqref{selfadjointfactor}
is a linear polynomial in $\D{i}$ of degree $p$. 
We easily find that if it is decomposed into separate equations
given by the coefficients of the polynomial, 
then the resulting equations are 
the same as 
the determining equations \eqref{splitsys}.

\Cor\label{thmvaradsymm}
Multipliers for any first-order \CK/ PDE system 
are completely characterized as 
adjoint symmetries with a variational form. 
\endCor

Moreover, it is interesting to note that 
the determining equations \eqref{splitsys} 
take the same form regardless of $\g{\sigma}$
for all first-order \CK/ PDE systems \eqref{uPDE}.

\subsection{Conservation law construction formula}
\label{formula}

We now give an integral formula that constructs 
the conserved densities $\P{t}$ and $\P{i}$ 
for any nontrivial \conslaw/ in normal form \eqref{normalcons}
in terms of its multipliers $\{ \factor{\sigma} \}$.
The formula makes use of the identities \cite{prlpaper}
\EQs
&& 
\W{\sigma} \linop{g}{\sigma}{\rho} \V{\rho} 
-\V{\rho} \adlinop{g}{\sigma}{\rho} \W{\sigma}
=\D{i} \S{i}[\V{},\W{};\g{}], 
\label{lingid}\\
&&
\Wt{\sigma} \linop{\factor{}}{}{\sigma\rho} \V{\rho}
-\V{\rho} \adlinop{\factor{}}{}{\sigma\rho} \Wt{\sigma}
=\D{i} \S{i}[\V{},\Wt{};\factor{}], 
\label{linfactorid}
\endEQs
where 
\EQs
&& 
\S{i}[\V{},\W{};\g{}] = 
\sum_{\ell=0}^{m-1} \sum_{k=0}^{m-\ell-1} 
(-1)^k ( \D{i_1}\cdots\D{i_\ell}\V{\rho} )
\D{j_1}\cdots\D{j_k}\Big( \W{\sigma} 
\Uderg{\sigma}{\rho}{i i_1\cdots i_k j_1\cdots j_\ell} \Big), 
\nonumber\\&&
\label{Sg}\\
&& 
\S{i}[\V{},\Wt{};\factor{}] = 
\sum_{\ell=0}^{p-1} \sum_{k=0}^{p-\ell-1} 
(-1)^k ( \D{i_1}\cdots\D{i_\ell}\V{\rho} )
\D{j_1}\cdots\D{j_k}\Big( \Wt{\sigma} 
\Uderfactor{\sigma}{\rho}{i i_1\cdots i_k j_1\cdots j_\ell} \Big), 
\nonumber\\&&
\label{Sfactor}
\endEQs
which are trilinear expressions 
derived by manipulation of the linearization operators and adjoint operators.
(Note, the terms in \Eqrefs{Sg}{Sfactor} with $\ell=0$ or $k=0$
are understood to involve no derivatives of $\V{}$ and $\W{}$, respectively.)

To set up the formula,
we first let
\EQ
\uparm{\sigma}{\lambda} =\lambda \u{\sigma} +(1-\lambda)\tu{\sigma}
\endEQ
where $\{ \tu{\sigma} \}$ are any functions of $t,\bdx$.
This defines a one-parameter $\lambda$ family of functions
with $\uparm{\sigma}{1}=\u{\sigma}$ and $\uparm{\sigma}{0}=\tu{\sigma}$.
Then we let
\EQs
&& 
\factor{\rho}[\bdul] =
\factor{\rho}(t,\bdx,\bdul,\der{\bdx}\bdul,\ldots,\nder{\bdx}{p}\bdul) , 
\\&&
\g{\rho}[\bdul] =
\g{\rho}(t,\bdx,\bdul,\der{\bdx}\bdul,\ldots,\nder{\bdx}{m}\bdul) , 
\\&&
\K(t,\bdx) = 
\Big( (\uparmder{\rho}{\lambda}{t} + \g{\rho}[\bdul]) \factor{\rho}[\bdul] 
\Big)\Bigm|_{\lambda=0} . 
\endEQs

\Thm\label{thmformula}
For the \CK/ PDE system \eqref{uPDE},
the conserved densities of any nontrivial conservation law
in normal form are given in terms of the multipliers by
\EQs
&& 
\P{t} = 
\int_0^1 d\lambda  (\u{\sigma}-\tu{\sigma}) \factor{\sigma}[\bdul] 
+ t\int_0^1 d\lambda  \K(\lambda t,\lambda \bdx)  , 
\label{Pt}\\
&&
\P{i} =
\x{i}\int_0^1 d\lambda \lambda^n \K(\lambda t,\lambda \bdx) 
+ \int_0^1 d\lambda  \Big( 
\S{i}[ \bdu-\bdtu,\bdfactor[\bdul];\bdg[\bdul] ]
\nonumber\\&&\fewquad 
+\S{i}[ \bdu-\bdtu,\bdg[\bdul]-\lambda\bdg[\bdu] +(1-\lambda)\bdDtu{t};
\bdfactor[\bdul] ] 
\Big) . 
\label{Pi}
\endEQs 
\endThm

In applying the construction formula \eqrefs{Pt}{Pi},
we must fix a choice for the functions $\{ \tu{\sigma} \}$.
If the expressions $\factor{\sigma}$ and $\g{\sigma}$
are nonsingular for $\u{\sigma}=0$,
then we can choose $\tu{\sigma}=0$ 
and this simplifies the integrals.
Moreover, if $\tu{\sigma}=\u{\sigma}=0$ 
satisfies the PDE system \eqref{uPDE},
then the $\K$ integrals vanish.

In the case when the expressions $\factor{\sigma}$ and $\g{\sigma}$
are singular at $\u{\rho}=0$ (for some $\rho=1,\ldots,N$),
we must choose $\tu{\rho}\neq 0$ such that
the expressions 
$\factor{\sigma}[\bdtu]$ and $\g{\sigma}[\bdtu]$
are nonsingular. 
It is sufficient to fix a simple choice of $\tu{\rho}$ 
such that the integrals converge.
Any change in the choice of $\tu{\rho}$ changes the conserved densities
only by a trivial conserved density \eqref{freedom}.

A simple proof of Theorem~\ref{thmformula} is given in \secref{proofs}.

\subsection{Proofs of Main Equations}
\label{proofs}

Recall that, for first-order \CK/ PDE systems \eqref{uPDE},
the proof of the determining system \eqref{splitsys}
for \conslaw/ multipliers in Theorem~\ref{thmdeteq}
reduces, by Lemma~\ref{lemsplit}, 
to the determining condition \eqref{multiplierdeteq} 
involving the Euler operator. 
To conclude this section, 
we present a simple, direct proof of 
this determining condition \eqref{multiplierdeteq} 
together with the construction formula \eqrefs{Pt}{Pi} 
for corresponding conserved densities in Theorem~\ref{thmformula}. 
The proof of \Eqref{multiplierdeteq} 
is based on an identity for linearization of 
the multiplier equation \eqref{multipliereq}.

We let 
\EQ
\uparm{\sigma}{\lambda}=(\lambda -1)\v{\sigma} + \u{\sigma}
\endEQ
be a one-parameter family of functions with 
$\uparm{\sigma}{1}=\u{\sigma}$
being an arbitrary function,
and with 
$\Parder{\uparm{\sigma}{\lambda}}{\lambda}=\v{\sigma}$
for any functions $\v{\sigma}(t,\bdx)$.

\Prop\label{propid}
For any given expressions 
$\factor{\sigma}[\bdu]= 
\factor{\sigma}(t,\bdx,\bdu,\der{\bdx}\bdu,\ldots,\nder{\bdx}{p}\bdu)$, 
$\Phat{t}[\bdu]= 
\Phat{t}(t,\bdx,\bdu,\der{\bdx}\bdu,\ldots,\nder{\bdx}{k}\bdu)$
and 
$\Phat{i}[\bdu]= 
\Phat{i}(t,\bdx,\bdu,\der{\bdx}\bdu,\ldots,\nder{\bdx}{k}\bdu)$, 
the following identities hold by direct calculation:
\EQs
(i) && 
\Parder{}{\lambda}\Big( 
(\uparmder{\sigma}{\lambda}{t} + \g{\sigma}[\bdul]) \factor{\sigma}[\bdul]
\Big)
\nonumber\\&&
= (\vder{\sigma}{t} +\linop{\bdg[\bdul]}{\sigma}{\rho} \v{\rho})
\factor{\sigma}[\bdul]
+ (\uparmder{\sigma}{\lambda}{t} + \g{\sigma}[\bdul])
\linop{\bdfactor[\bdul]}{}{\sigma\rho} \v{\rho}
\nonumber\\&&
= \v{\sigma} \Big(
-\D{t}\factor{\sigma}[\bdul]
+\adlinop{\bdg[\bdul]}{\rho}{\sigma}\factor{\rho}[\bdul]
+\adlinop{\bdfactor[\bdul]}{}{\sigma\rho}
(\uparmder{\rho}{\lambda}{t} + \g{\rho}[\bdul])
\Big)
\nonumber\\&&\qquad 
+\D{t}\Big( \v{\sigma}\factor{\sigma}[\bdul] \Big)
+\D{i}\Big( \S{i}[ \bdv,\bdfactor[\bdul];\bdg[\bdul] ] 
\nonumber\\&&\fewquad 
+\S{i}[ \bdv,\bdDul{t}+\bdg[\bdul];\bdfactor[\bdul] ] \Big)
\label{linRS}
\endEQs
where $\S{i}$ denotes the trilinear expressions given by \Eqrefs{Sg}{Sfactor};
\EQ
(ii) 
\Parder{}{\lambda}\Big( 
\D{t}\Phat{t}[\bdul] +\D{i}\Phat{i}[\bdul] \Big) =
\D{t}( \linop{\Phat{t}[\bdul]}{}{\sigma} \v{\sigma} )
+\D{i}( \linop{\Phat{}[\bdul]}{i}{\sigma} \v{\sigma} )
\label{linLS}
\endEQ
where
$\linop{\Phat{t}}{}{\sigma}$
and $\linop{\Phat{}}{i}{\sigma}$ 
denote the linearization operators of $\Phat{t}$ and $\Phat{i}$ respectively. 
\endProp

\Proof[Proof of the multiplier determining condition
and conserved density construction formula:]

Suppose $\P{t},\P{i}$ are conserved densities of a \conslaw/ 
in normal form \eqref{normalcons}. 
From Theorem~\ref{thmmult} 
the multipliers for the \conslaw/ are given by 
$\factor{\sigma}=\ELophat{u}{\sigma} (\P{t})$
satisfying the multiplier equation \eqref{multipliereq}
with $\Phat{t}= \P{t}, \Phat{i}= \P{i}-\Gamma^i$. 

Since the multiplier equation \eqref{multipliereq} 
holds for all functions $\u{\sigma}(t,\bdx)$,
it must hold for the one-parameter family $\uparm{\sigma}{\lambda}$.
We now take the derivative of 
the resulting left-side and right-side expressions of \Eqref{multipliereq}
with respect to $\lambda$. 
By Proposition~\ref{propid}, on the left-side we obtain \Eqref{linRS},
while on the right-side we directly obtain \Eqref{linLS}. 
These expressions \eqrefs{linRS}{linLS} 
are equal for all functions $\v{\sigma}(t,\bdx)$
and therefore hold iff the terms multiplying $\v{\sigma}$ vanish
and the total derivative terms involving $\v{\sigma}$ are separately equal
(by considering the terms $\vder{\sigma}{t}, \vder{\sigma}{i}$). 
From the terms multiplying $\v{\sigma}$ we have 
\EQ\label{Ufactoreq}
0=
-\D{t}\factor{\sigma}[\bdul]
+\adlinop{\bdg[\bdul]}{\rho}{\sigma}\factor{\rho}[\bdul]
+\adlinop{\bdfactor[\bdul]}{}{\sigma\rho}
(\uparmder{\rho}{\lambda}{t} + \g{\rho}[\bdul]) . 
\endEQ
This reduces when $\lambda=1$ 
to \Eqref{multiplierdeteq}
and hence $\{ \factor{\sigma} \}$ is a solution of
the determining condition \eqref{multiplierdeteq}.

Conversely, suppose $\{ \factor{\sigma} \}$ is a solution of 
the determining condition \eqref{multiplierdeteq}.
Then, by combining the two identities in Proposition~\ref{propid}, 
we see 
$\factor{\sigma}$ satisfies the linearized multiplier equation
\EQ
\Parder{}{\lambda}\Big(
(\uparmder{\sigma}{\lambda}{t} 
+ \g{\sigma}[\bdul]) \factor{\sigma}[\bdul] \Big)
= \D{t} \Parder{}{\lambda}\Phat{t}[\bdul] 
+\D{i} \Parder{}{\lambda}\Phat{i}[\bdul]
\endEQ
with $\Parder{\Phat{t}[\bdul]}{\lambda}$
and $\Parder{\Phat{i}[\bdul]}{\lambda}$
defined by 
\EQs
&& 
\linop{\Phat{t}[\bdul]}{}{\sigma} \v{\sigma}
=\v{\sigma}\factor{\sigma}[\bdul]
+\D{i} \theta^i , 
\label{Ptlinid}
\\&&
\linop{\Phat{}[\bdul]}{i}{\sigma} \v{\sigma}
=( \S{i}[ \bdv,\bdfactor[\bdul];\bdg[\bdul] ]
+\S{i}[ \bdv,\bdDul{t}+ \bdg[\bdul];\bdfactor[\bdul] ] )
-\D{t}\theta^i +\D{j}\psi^{ij} , 
\nonumber\\&&
\label{Pilinid}
\endEQs
for some expressions $\theta^i,\psi^{ij}=-\psi^{ji}$.
We now undo the linearization to obtain 
the multiplier equation \eqref{multipliereq}
by integrating with respect to $\lambda$ as follows.
We set $\v{\sigma}=\u{\sigma}-\tu{\sigma}$, and so
\EQ
\uparm{\sigma}{\lambda} = \lambda (\u{\sigma}-\tu{\sigma}) +\tu{\sigma} . 
\endEQ
Then we use the fundamental theorem of calculus to obtain
\EQ\label{FTC}
(\uder{\sigma}{t} +\g{\sigma}[\bdu]) \factor{\sigma}[\bdu]
= \D{t} \Phat{t}[\bdu] +\D{i} \Phat{i}[\bdu]
+(\tuder{\sigma}{t} +\g{\sigma}[\bdtu]) \factor{\sigma}[\bdtu]
-\D{t}\Phat{t}[\bdtu] -\D{i}\Phat{i}[\bdtu]
\endEQ
where 
\EQs
\Phat{t}[\bdu] = && 
\Phat{t}[\bdtu] 
+ \int_0^1 d\lambda (\u{\sigma}-\tu{\sigma}) \factor{\sigma}[\bdul] , 
\label{Phatt}\\
\Phat{i}[\bdu] = && 
\Phat{i}[\bdtu]
+\int_0^1 d\lambda ( 
\S{i}[ \bdu-\bdtu,\bdfactor[\bdul];\bdg[\bdul] ]
\nonumber\\&&\fewquad
+\S{i}[ \bdu-\bdtu,\bdDul{t}+\bdg[\bdul];\bdfactor[\bdul] ] ) , 
\label{Phati}
\endEQs
to within trivial conserved densities.
Since \Eqref{FTC} holds for all $u(t,\bdx)$,
while $\tu{}(t,\bdx)$ is fixed, 
we must have 
\EQ\label{Keq}
\D{t}\Phat{t}[\bdtu] +\D{i}\Phat{i}[\bdtu]
= (\tuder{\sigma}{t} +\g{\sigma}[\bdtu]) \factor{\sigma}[\bdtu]
= \K(t,\bdx) . 
\endEQ
It is then simple to check that \Eqref{Keq} is satisfied identically by
\EQ\label{Kterms}
\Phat{t}[\bdtu] = t\int_0^1 d\lambda \K(\lambda t,\lambda \bdx) ,
\Phat{i}[\bdtu] = \x{i} \int_0^1 d\lambda \K(\lambda t,\lambda \bdx) . 
\endEQ

Thus, we find from \Eqref{FTC} that
$\{ \factor{\sigma} \}$ satisfies 
the multiplier equation \eqref{multipliereq}, 
with conserved densities given by \Eqsref{Phatt}{Kterms}.
Hence, by Theorem~\ref{thmmult}, 
$\factor{\sigma}$ are multipliers for a \conslaw/ 
in normal form \eqref{normalcons}. 

To obtain the construction formula \eqrefs{Pt}{Pi}
for the conserved densities,
we move onto the solution space of \Eqref{uPDE}
and substitute 
$\uparmder{\sigma}{\lambda}{t} 
= -\lambda\g{\sigma}[\bdu] +(1-\lambda) \tuder{\sigma}{t}$
into \Eqrefs{Phatt}{Phati}. 
The expressions $\Phat{t}[\bdu]$ and $\Phat{i}[\bdu]$
directly reduce to the formula for $\P{t}$ and $\P{i}$.
\endProof

\section{Treatment of Nth order scalar PDEs}
\label{scalarcase}

Here we exhibit the \conslaw/ determining system and construction formula
for scalar PDEs of any order 
with one dependent variable $u$ 
and $n+1$ independent variables $t,\bdx=(\x{1},\ldots,\x{n})$.
We work directly with the scalar PDE expressed in 
an $N$th order \CK/ form
\EQ\label{scalarPDE}
\G{}= \purederu{}{\N}{t} +\g{}(t,x,u,\ujet{},\ldots,\ujet{m}) =0
\endEQ
where in this section
$\ujet{q}$ now denotes all derivatives of $u$ of order $q$,
excluding $t$ derivatives of $u$ of order $q\ge N$ 
and their differential consequences
(\ie/ the PDE is written so that 
the $t$ derivatives of $u$ of highest order appear in solved form).

Clearly, without loss of generality, 
for \conslaw/s 
we are free to eliminate $\N$th order $t$ derivatives of $u$ 
(and differential consequences) 
in considering conserved densities. 

\Defn\label{defnscalarCK}
A {\it local \conslaw/ in normal form} 
for a \CK/ scalar PDE \eqref{scalarPDE} 
is a divergence expression
\EQ\label{cons}
\D{t}\P{t}(t,\bdx,u,\ujet{},\ldots,\ujet{k}) 
+ \D{i}\P{i}(t,\bdx,u,\ujet{},\ldots,\ujet{k}) = 0
\endEQ
holding for all solutions $u(t,\bdx)$ of \Eqref{scalarPDE}. 
\endDefn

A \conslaw/ \eqref{cons} is trivial 
if it holds as an identity \eqref{trivialcons}
for some expressions 
$\theta^i(t,\bdx,u,\ujet{},\ldots,\ujet{k-1})$,
$\psi^{ij}(t,\bdx,u,\ujet{},\ldots,\ujet{k-1})$
with $\psi^{ij}=-\psi^{ji}$,  
for all solutions $u(t,\bdx)$ of PDE \eqref{scalarPDE}.
Only nontrivial \conslaw/s \eqref{cons} are of interest.

All nontrivial \conslaw/s \eqref{cons} of PDE \eqref{scalarPDE}
can be shown to arise from
multipliers on the PDE, 
similarly to Theorem~\ref{thmmult}. 
We move off the solution space of \Eqref{scalarPDE}
and let $u(t,\bdx)$ be an arbitrary function of $t,\bdx$.
We use the notation
$\nder{t}{q} u= \purederu{}{q}{t}$ 
for pure $t$ derivatives of $u$,
and $\Du{i}=\purederu{}{}{\x{i}}$, 
$\Du{ij}=\mixderu{}{2}{\x{i}}{\x{j}}$, \etc/
for pure $\bdx$ derivatives of $u$,
and $\nder{t}{q}\Du{i}= \mixderu{}{q+1}{t^q}{\x{i}}$, 
$\nder{t}{q}\Du{ij}= \mixderu{}{q+2}{t^q}{\x{i}\partial\x{j}}$, \etc/ 
for mixed $t,\bdx$ derivatives of $u$,
with $\nder{0}{q} u= u$ and $\nder{t}{0}\Du{i}= \Du{i}$. 

\Thm\label{thmscalarmult}
For the \CK/ scalar PDE \eqref{scalarPDE}, 
{\rm every} nontrivial \conslaw/ \eqref{cons} 
is uniquely characterized by a multiplier $\factor{}$ 
with no dependence on $\nderu{}{t}{\N}$ and differential consequences.
The multiplier satisfies the relations 
\EQ\label{scalarconseq}
( \nderu{}{t}{\N} +\g{} ) \factor{} 
=\D{t}\P{t} +\D{i}(\P{i}-\Gamma^i)
\endEQ
and 
\EQ\label{scalarfactoreq}
\factor{} = \ELophat{}{} (\P{t})
\endEQ
holding for all functions $u(t,\bdx)$,
where 
\EQ
\ELophat{}{} = 
\Parder{}{(\nder{t}{\N-1}u)} -\D{i}\Parder{}{(\nder{t}{\N-1}\Du{i})}
+\D{i}\D{j}\Parder{}{(\nder{t}{\N-1}\Du{ij})} 
+\cdots
\endEQ
is a restricted Euler operator,
and $\Gamma^i$ is given by an expression proportional to 
$\nderu{}{t}{\N}+\g{}$
and its differential consequences. 
\endThm
From \Eqref{scalarfactoreq}
one can show that $\factor{}$ is invariant under a change in $\P{t}$ by
a trivial conserved density \eqref{trivialcons}. 
(In particular, if $\P{t}$ is trivial, 
then $\factor{}$ is identically zero, 
and conversely.)
Consequently, it is natural to define 
the order of a \conslaw/ \eqref{cons} 
as the order of the highest derivatives of $u$ 
in its multiplier \eqref{scalarfactoreq}.

It is straightforward to derive both
the determining system for multipliers $\factor{}$
and the construction formula for conserved densities in terms of $\factor{}$
by applying the results in \secrefs{multiplierdetsys}{formula}
to the scalar PDE \eqref{scalarPDE}
written as a first-order \CK/ system
(which we carry out later). 

In order to display the determining equations explicitly,
we introduce the $\N+1$ expressions
\EQs
\Omega_0 = && \factor{}, 
\nonumber\\
\Omega_q = && (-1)^q \solnD{q}{t} \factor{}
+\sum_{k=1}^q (-1)^{q-k} \solnD{q-k}{t} \bigg(
\parderg{}{(\nder{t}{\N-k}u)} \factor{}
- \D{i}\Big( \parderg{}{(\nder{t}{\N-k}\Du{i})} \factor{} \Big)
+\cdots
\nonumber\\&&\fewquad
+ (-1)^m \D{i_1}\cdots \D{i_m}\Big(
\parderg{}{(\nder{t}{\N-k}\Du{i_1\cdots i_m})} \factor{} \Big)
\bigg),\
q=1,\ldots,\N
\nonumber\\
\label{sysfactors}
\endEQs
where $\solD{t}$ is the total derivative operator with respect to $t$ 
on the solution space of the PDE \eqref{scalarPDE}
as defined by eliminating 
$\nderu{}{t}{\N}=-g$ and all differential consequences.
(In particular, 
$\solD{t} u=\der{t}u$, $\solnD{2}{t}u =\nder{t}{2}u$, \etc/,
and $\solnD{\N}{t} u= -g$.)
Note that, 
if the order of $\Omega_0$ with respect to $\bdx$ derivatives of $u$ is $p$,
the order of $\Omega_q$ is at most $p+mq$. 

\Thm\label{thmscalardeteq}
For the \CK/ scalar PDE \eqref{scalarPDE},
the multipliers for {\rm all} nontrivial \conslaw/s \eqref{cons}
up to any given order $p$
are the solutions \break
$\factor{}(t,\bdx,u,\ujet{},\ldots,\ujet{p})$
of the determining system 
\EQ\label{scalaradsymmsys}
\Omega_\N =0 
\endEQ
and
\EQs
&&
\Parder{\Omega_{k}}{(\nder{t}{j}u)} 
-\Parder{\Omega_{j}}{(\nder{t}{k}u)} 
= \sum_{k=1}^{p'} (-1)^k \D{i_1} \cdots \D{i_k} 
\Parder{\Omega_{k}}{(\nder{t}{j}\Du{i_1\cdots i_k})} , 
\nonumber\\
&&
\Parder{\Omega_{k}}{(\nder{t}{j}\Du{i_1\cdots i_q})} 
-(-1)^q \Parder{\Omega_{j}}{(\nder{t}{k}\Du{i_1\cdots i_q})} 
\nonumber\\&&\qquad
= \sum_{k=q+1}^{p'} (-1)^{k-q+1} \frac{k!}{q!(k-q)!} 
\D{i_{q+1}} \cdots \D{i_{k}} 
\Parder{\Omega_{k}}{(\nder{t}{j}\Du{i_1\cdots i_{k}})} , 
q=1,\ldots,{p'-1}
\nonumber\\
&&
\Parder{\Omega_{k}}{(\nder{t}{j}\Du{i_1\cdots i_{p'}})} 
- (-1)^{p'} \Parder{\Omega_{j}}{(\nder{t}{k}\Du{i_1\cdots i_{p'}})}
=0, 
\nonumber\\&&
\label{scalarextrasys}
\endEQs 
where $p'=p+mk$, $j=0,1,\ldots,\N-1$; $k=0,1,\ldots,\N-1$. 
\endThm

In this system, 
\Eqref{scalaradsymmsys} is the determining equation 
for the adjoint symmetries 
$\factor{} = \adsymm{}(t,\bdx,u,\ujet{},\ldots,\ujet{p})$ 
of order $p$ 
of the PDE \eqref{scalarPDE}, 
explicitly 
\EQ\label{scalaradsymmeq}
0=(-\solD{t})^\N\adsymm{} +\adslinop{g} \adsymm{} . 
\endEQ
The extra determining equations \eqref{scalarextrasys}
are the necessary and sufficient conditions
for an adjoint symmetry to be a \conslaw/ multiplier. 
Since \Eqrefs{scalaradsymmsys}{scalarextrasys}
do not involve $\nderu{}{t}{\N}$ or any of its differential consequences, 
one is able to work equivalently on 
the solution space of the PDE \eqref{scalarPDE}
in order to find the solutions $\factor{}$.

In order now 
to display explicitly the construction formula for 
the conserved densities $\P{t},\P{i}$ 
in terms of the multiplier $\factor{}$, 
we first define the trilinear expression
\EQs
\S{i}[V,W;F] = &&
\sum_{j=0}^{\N-1} \bigg(
\Dt{j}V \Big( 
\Parder{F}{(\nder{t}{j}\Du{i})} W 
-\D{i_1} (\Parder{F}{(\nder{t}{j}\Du{ii_1})} W ) 
+\cdots \Big)
\nonumber\\&&
+\Dt{j} \D{j_1}V \Big( 
\Parder{F}{(\nder{t}{j}\Du{ij_1})} W  
-\D{i_1} (\Parder{F}{(\nder{t}{j}\Du{ii_1j_1})} W ) 
+\cdots \Big) +\cdots \bigg)
\label{trilin}
\endEQs
depending on arbitrary functions $V,W,F$. 
Next we let
\EQ
\uparm{}{\lambda} =\lambda \u{} +(1-\lambda)\tu{}
\endEQ
where $\tu{}$ is any function of $t,\bdx$.
This defines a one-parameter $\lambda$ family of functions
with $\uparm{}{1}=\u{}$ and $\uparm{}{0}=\tu{}$.
Then we define
\EQs
&& 
\Omega_q[\ul] =
\Omega_q(t,\bdx,\ul,\uljet{},\ldots,\uljet{p}),\
q=0,1,\ldots,\N-1
\\&&
\g{}[\ul] =
\g{}(t,\bdx,\ul,\uljet{},\ldots,\uljet{p}) , 
\\&&
\K(t,\bdx) = 
(\nder{t}{\N}\tu{}+ \g{}[\tu{}]) \Omega_0[\tu{}], 
\endEQs
using \Eqref{sysfactors} for $\Omega_q$ in terms of $\factor{}$.

\Thm\label{thmscalarformula}
For the \CK/ scalar PDE \eqref{scalarPDE},
the conserved densities of any nontrivial \conslaw/ \eqref{cons}
are given in terms of the multiplier $\factor{}$ by
\EQs
&& 
\P{t} = 
\int_0^1 d\lambda  \sum_{j=0}^{\N-1}
(\nderu{}{t}{j}-\nder{t}{j}\tu{}) \Omega_{j}[\ul] 
+ t\int_0^1 d\lambda  \K(\lambda t,\lambda \bdx)  , 
\label{scalarPt}\\
&&
\P{i} =
\x{i}\int_0^1 d\lambda \lambda^n \K(\lambda t,\lambda \bdx) 
+ \int_0^1 d\lambda  \Big( 
\S{i}[ \u{}-\tu{},\Omega_0[\ul];\g{}[\ul] ]
\nonumber\\&&\fewquad 
+\S{i}[ \u{}-\tu{},\g{}[\ul]-\lambda\g{}[u] +(1-\lambda)\nder{t}{\N}\tu{};
\Omega_0[\ul] ]
\Big) . 
\label{scalarPi}
\endEQs 
\endThm

In applying the construction formula \eqrefs{scalarPt}{scalarPi},
we fix the function $\tu{}$ 
so that the expressions 
$\factor{}[\tu{}]$ and $\g{}[\tu{}]$
are nonsingular. 
In particular, if $\factor{}[0]$ and $\g{}[0]$
are nonsingular 
then we can choose $\tu{}=0$,
which significantly simplifies the integrals.
Moreover, if $\tu{}=\u{}=0$ 
satisfies the PDE \eqref{scalarPDE},
then immediately the $\K$ integrals vanish.
A change in the choice of $\tu{}$ alters the conserved densities
only by a trivial conserved density \eqref{trivialcons}.

\Rem[Conversion to a first order \CK/ system:]

We now outline the proof of Theorems~\ref{thmscalardeteq} and~\ref{thmscalarformula}
using Theorems~\ref{thmdeteq} and~\ref{thmformula}.
To begin we write the scalar PDE \eqref{scalarPDE} 
in first-order (evolution) form \eqref{uPDE} with respect to $t$ 
as follows:
\EQs
&& 
\u{1}=u, 
\u{2}=\deru{}{t}, \ldots,
\u{\N}=\nderu{}{t}{\N-1}, 
\label{uident}\\
&& 
\g{1}=-\u{2}, \ldots,
\g{\N-1}=-\u{\N}, 
\g{\N}=\g{}, 
\label{gident}\\
&&
\G{1}=\der{t}\u{1} -\u{2} = 0, \ldots,
\G{\N-1}=\der{t}\u{\N-1} - \u{\N}=0, 
\G{\N}=\der{t}\u{\N} + \g{}=0. 
\label{scalarsys}
\endEQs
Through \Eqsref{uident}{scalarsys}
there is a one-to-one correspondence between
nontrivial \conslaw/s \eqref{cons} of the scalar PDE \eqref{scalarPDE}
and nontrivial \conslaw/s in normal form \eqref{normalcons}
of the equivalent first-order PDE system \eqref{scalarsys}.
The relation between 
a multiplier $\factor{}$ of a scalar PDE \conslaw/
and a set of multipliers $\{ \factor{1},\ldots,\factor{\N} \}$ 
of the corresponding PDE system \conslaw/
can be obtained by considering the adjoint symmetry equations
of the scalar PDE \eqref{scalarPDE} and the PDE system \eqref{scalarsys}.
Straightforwardly, from \Eqrefs{gident}{adsymmsys}, 
we have 
\EQs
&& 0= -\solD{t} \factor{\N} +\adslinop{0,g} \factor{1}, 
\label{scalarsysadsymmeq}\\
&& 0= -\solD{t} \factor{\N-q} -\factor{\N-q+1} 
+ \adslinop{q,g} \factor{1}, 
q=1,\ldots,\N-1 
\label{scalarsysadsymmeq'}
\endEQs
where 
$\adslinop{q,g}$ is the adjoint operator of 
the linearization operator $\slinop{q,g}$ defined by 
\EQ
\slinop{q,g} = 
\parderg{}{(\nder{t}{q}u)} + \parderg{}{(\nder{t}{q}\Du{i})}\D{i} + \cdots
+\parderg{}{(\nder{t}{q}\Du{i_1\cdots i_m})}\D{i_1}\cdots\D{i_m} . 
\endEQ
By solving \Eqref{scalarsysadsymmeq'} for 
$\factor{2},\ldots,\factor{\N}$ in terms of $\factor{1}$
and comparing \Eqref{scalarsysadsymmeq} with \Eqref{scalaradsymmsys}, 
we directly see 
\EQ\label{factorident}
\factor{1}=\factor{}=\Omega_0,
\factor{2}=\Omega_1, \ldots,
\factor{\N}=\Omega_{\N-1}. 
\endEQ
This establishes an explicit correspondence between
$\factor{}$ and $\{ \factor{1},\ldots,\factor{\N} \}$
leading immediately to Theorems~\ref{thmscalardeteq} and~\ref{thmscalarformula}
from Theorems~\ref{thmdeteq} and~\ref{thmformula}.

\Rem[Remarks on the determining system and construction formula:]

Theorems~\ref{thmscalardeteq} and~\ref{thmscalarformula} can also be established 
directly from Theorem~\ref{thmscalarmult}
without use of the results in \secrefs{multiplierdetsys}{formula}. 
The main step in the proof of Theorem~\ref{thmscalardeteq} 
is a polynomial splitting result
analogous to Lemma~\ref{lemsplit} as follows.

The determining condition for a multiplier $\factor{}$ of order $p$
for the scalar PDE \eqref{scalarPDE}
arises from the relation \eqref{scalarconseq}
by the result that an expression is a divergence if and only if 
it is annihilated by the full Euler operator
\EQ
\ELop{u}{} =
\Parder{}{u} -\D{i}\Parder{}{\Du{i}} -\D{t}\Parder{}{(\der{t}u)} 
+\D{i}\D{j}\Parder{}{\Du{ij}} 
+\D{t}\D{j}\Parder{}{(\der{t}\Du{j})} 
+\nD{2}{t}\Parder{}{(\nder{t}{2}u)}
+\cdots . 
\endEQ
This can be shown (by a straightforward calculation \cite{olverbook}) 
to yield
\EQ\label{scalarmultiplierdeteq}
0=\ELop{u}{} ( (\nderu{}{t}{\N})\factor{} + \g{}\factor{} )
= (-\D{t})^\N \factor{} +\adslinop{g}\factor{}
+\adslinop{\factor{}} (\nderu{}{t}{\N} +\g{}) , 
\endEQ
which is required to hold for all functions $u(t,\bdx)$ 
(not just solutions of \Eqref{scalarPDE}). 
The determining condition \eqref{scalarmultiplierdeteq}
is a polynomial in 
$\nderu{}{t}{\N},\nderu{}{t}{\N+1},\ldots,\nderu{}{t}{2\N-1}$
and differential consequences with respect to $\bdx$. 
Furthermore, the terms in this polynomial have weights $0$ up to $\N$,
where we assign weight 
$1$ to $\nderu{}{t}{\N}$ (and $\bdx$ derivatives of $\nderu{}{t}{\N}$),
$2$ to $\nderu{}{t}{\N+1}$ (and $\bdx$ derivatives of $\nderu{}{t}{\N+1}$), 
\etc/,
and we add the weights of products (and powers) of 
$\nderu{}{t}{\N},\nderu{}{t}{\N+1}$, \etc/.
Now, since $u$ is required to be an arbitrary function of $t$ and $\bdx$, 
the polynomial splits into separate determining equations
given by the coefficients of the various weight terms
involving $\nderu{}{t}{\N},\nderu{}{t}{\N+1},\ldots,\nderu{}{t}{2\N-1}$
(and differential consequences with respect to $\bdx$).
It is convenient to organize the splitting by working in terms of
$\nder{t}{\N}u+\g{}=\G{}$, 
$\nder{t}{\N+1}u+\solD{t}\g{}=\solD{t}\G{}$, 
$\nder{t}{\N}\Du{i}+\D{i}\g{}=\D{i}\G{}$, 
$\nder{t}{\N+1}\Du{i}+\D{i}\solD{t}\g{}=\D{i}\solD{t}\G{}$, \etc/.
The terms of weight $0$ yield 
the adjoint symmetry determining equation \eqref{scalaradsymmsys}
and the terms of weight $1$ up to $\N$ yield
the extra determining equations \eqref{scalarextrasys}
on $\factor{}$. 
This derivation is illustrated in the second example of Part I.

The construction formula for conserved densities $\P{t}$ and $\P{i}$
of a \conslaw/ for PDE \eqref{scalarPDE}
is obtained 
by inverting the Euler operator equation \eqref{scalarmultiplierdeteq}
as follows. 
Since \Eqref{scalarmultiplierdeteq} holds for arbitrary functions $u(t,\bdx)$,
it must hold with $u$ replaced by the one-parameter family 
$\uparm{}{\lambda} =\lambda \u{} +(1-\lambda)\tu{}$.
This yields
\EQ\label{linscalardeteq}
0= (-\D{t})^\N \factor{}[\ul] + \adslinop{\g{}[\ul]} \factor{}[\ul]
+ \adslinop{\factor{}[\ul]} ( \nder{t}{\N}\uparm{}{\lambda} +\g{}[\ul] ) . 
\endEQ
We multiply \Eqref{linscalardeteq} by $\u{}-\tu{}$
and then rearrange the terms which involve total derivative operators
coming from $\adslinop{\g{}}$ and $\adslinop{\factor{}}$.
This leads to the formula
\EQs
&& 
\D{t}\Big(  
\sum_{j=0}^{\N-1} (\nder{t}{j}u-\nder{t}{j}\tu{}) \Omega_{j}[\ul] \Big)
+\D{i}\Big( 
\S{i}[ \u{}-\tu{},\Omega_0[\ul];\g{}[\ul] ]
\nonumber\\&&\fewquad
+\S{i}[ \u{}-\tu{},\nder{t}{\N}\uparm{}{\lambda} + \g{}[\ul];\Omega_0[\ul] ]
\Big)
= \Parder{}{\lambda}\Big( 
(\nder{t}{\N}\uparm{}{\lambda} +\g{}[\ul]) \factor{}[\ul] \Big) . 
\endEQs
Next we integrate from $\lambda=0$ to $\lambda=1$ 
and apply the fundamental theorem of calculus.
Using the identity 
$\D{t}\Big( t\int_0^1 d\lambda  \K(\lambda t,\lambda \bdx) \Big)
+\D{i}\Big( \x{i}\int_0^1 d\lambda \lambda^n \K(\lambda t,\lambda \bdx) \Big)
=\K$,
and finally moving onto the solution space of the PDE \eqref{scalarPDE},
we obtain the \conslaw/ \eqref{cons}
with $\P{t}$ and $\P{i}$ given by \Eqrefs{scalarPt}{scalarPi}.

\Rem[Remarks on variational principles:]

\Defn\label{defnvarPDE}
A \CK/ scalar PDE \eqref{scalarPDE} is called {\it variational}
if it arises from an action 
\EQ\label{action}
S=\int\Big( 
L(t,\bdx,u,\ujet{},\ldots,\ujet{k}) 
\Big) dt d\bdx
\endEQ
by variation with respect to $u$, 
\EQ
G=\ELop{u}{}(L) = \nderu{}{t}{\N} +\g{}. 
\endEQ 
\endDefn

The well-known necessary and sufficient condition \cite{olverbook}
for existence of an action \eqref{action} is that
\EQ
\nD{\N}{t} + \slinop{\g{}} = (-\D{t})^\N +\adslinop{\g{}} , 
\endEQ
\ie/ $\N$ must be even and $\g{}$ must have a self-adjoint linearization. 
This condition is equivalent to requiring that 
the determining equation for symmetries of the PDE \eqref{scalarPDE}
is self-adjoint.

In the case when PDE \eqref{scalarPDE} is variational, 
Theorem~\ref{thmscalardeteq} combined with Noether's theorem 
\cite{blumanbook,olverbook}
shows that the extra determining equations \eqref{scalarextrasys}
constitute necessary and sufficient conditions for 
a symmetry of the PDE \eqref{scalarPDE} 
to leave invariant the action \eqref{action} 
to within a boundary term. 
In particular, if 
$\X{}u = \symm{}(t,\bdx,u,\ujet{},\ldots,\ujet{p})$ 
is a symmetry of order $p$, 
then 
$\X{}S=\int( \D{t}\theta^t +\D{i}\theta^i ) dt d\bdx$
holds for some expressions $\theta^t$ and $\theta^i$ iff
$\factor{}= \symm{}$ satisfies \Eqref{scalarextrasys}
and hence $\symm{}$ is a multiplier yielding a \conslaw/ \eqref{cons}
of PDE \eqref{scalarPDE}.

\section{Summary and concluding remarks}
\label{conclude}

For any \CK/ system $G$ of one or more PDEs, 
Theorems~\ref{thmdeteq},~\ref{thmformula} 
and Theorems~\ref{thmscalardeteq},~\ref{thmscalarformula}
yield an effective computational method
to obtain all local \conslaw/s (up to any specified order). 
The method is summarized as follows:

1. Linearize $G$ to form its linearized system $\ell$,
which is the determining system for the symmetries of $G$.

2. Form the \adsys/ $\ell^*$ of $\ell$,
which is the determining system for the adjoint symmetries of $G$.

3. Form the \hsys/ $\h$ comprising
the necessary and sufficient determining equations 
for an adjoint symmetry to be a multiplier for a \conslaw/ of $G$. 

4. Solve the augmented system $\ell^* \cup \h$. 
This is the determining system for the multipliers that yield 
all nontrivial local \conslaw/s of $G$. 

5. Use the explicit construction formula to obtain 
the conserved densities arising for each solution of the system 
$\ell^* \cup \h$. 

The linearized system of $G$ is self-adjoint ($\ell = \ell^*$)
if and only if $G$ is variational,
in which case 
solutions of $\ell^*$ are solutions of $\ell$. 
Then the \hsys/ $\h$ is equivalent to the condition for symmetries 
to leave invariant the action for $G$. 
In general, if $G$ is not variational 
then 
solutions of $\ell^*$ are not solutions of $\ell$.
 
The systems $\ell,\ell^*,\h$, and $\ell^* \cup \h$
are all linear overdetermined systems 
which are solved working entirely
on the space of solutions of $G$ 
(\ie/ a leading derivative of the dependent variables in $G$ is eliminated). 
There exist algorithmic procedures \cite{software} 
to seek solutions of $\ell$.
These procedures can be readily adapted 
for seeking solutions of 
$\ell^*$, $\h$, and $\ell^* \cup \h$. 
In general, $\ell^* \cup \h$ is more overdetermined than $\ell$
and hence is typically easier to solve. 
More significantly, 
one can choose appropriate mixings of the determining equations 
in $\ell^*$ and $\h$ to solve $\ell^* \cup \h$ effectively.

One can also use specific ansatze to seek particular solutions of 
$\ell^* \cup \h$,
such as restricting the form of highest derivatives 
of the dependent variables of $G$ 
allowed in the solution. 
For example, familiar \conslaw/s such as energy invariably arise 
from the simple ansatz of seeking multipliers 
restricted to be linear in first derivatives.

In general it is important to note that
solutions of $\ell^*$ are not necessarily solutions of $\h$
and hence $\ell^*$ does not determine a \conslaw/ multiplier.
This typically occurs for
scaling symmetries of systems $G$ in the case $\ell^*=\ell$
(\ie/ self-adjoint),
and for point-type adjoint symmetries 
(first-order and linear in derivatives of dependent variables)
of systems $G$ in the case $\ell^*=-\ell$ 
(\ie/ skew-adjoint).
Examples are $u_{tt} -u_{xx} +u^3=0$
which has $u+tu_t+xu_x$ as a solution of $\ell=\ell^*$ 
but not a solution of $\h$;
$u_t+u_{xxx}=0$ which has $u_x$ as a solution of $\ell^*=-\ell$ 
but not a solution of $\h$. 
\Ref{odepaper} exhibits several ODE examples 
in which nontrivial adjoint symmetries are not multipliers. 
The need for the extra conditions $\h$ to determine multipliers
has not been clearly recognized in the literature
(\eg/ \cite{incomplete}).

The chief aspect of our method 
compared to other existing treatments of PDE \conslaw/s
(\eg/ \cite{fokas,vinogradov,olverbook,blumanbook,wolf})
is the explicit delineation of 
the linear determining system $\ell^* \cup \h$
which incorporates (and identifies) the necessary and sufficient conditions
for adjoint symmetries to be multipliers,
without moving off the space of solutions of the given PDE(s) $G$.
Consequently, one can calculate multipliers of \conslaw/s 
by effective algorithmic procedures.
Moreover there is the added computational advantage of 
allowing the determining equations in 
the adjoint system $\ell^*$ and the extra system $\h$ to be mingled
to optimally solve the determining system $\ell^* \cup \h$,
as illustrated by the \conslaw/ classification results
for the PDE examples in Part I.

\begin{acknowledgments}
The authors are supported in part by the
Natural Sciences and Engineering Research Council of Canada.
We gratefully thank the referees for useful comments
which have improved this paper. 
\end{acknowledgments}

\end{document}